# Wavelength agile multi-photon microscopy with a fiber amplified diode laser

**Matthias Eibl[1], Daniel Weng[1], Hubertus Hakert[1], Jan Philip Kolb[1], Tom Pfeiffer[1], Jennifer E. Hundt[2], Robert Huber[1], and Sebastian Karpf [1,\*]**

[1]*Institute of Biomedical Optics, University of Lübeck, 23562 Lübeck, Germany*
[2]*Lübeck Institute of Experimental Dermatology (LIED), University of Lübeck, 23538 Lübeck, Germany*
*\*karpf@bmo.uni-luebeck.de*

**Abstract:** Multi-photon microscopy is a powerful tool in biomolecular research. Less complex and more cost effective excitation light sources will make this technique accessible to a broader community. Especially semiconductor diode seeded fiber lasers have proven to be robust, low cost and easy to use. However, their wavelength tuning range is often limited, so only a limited number of fluorophores can be accessed. Therefore, different approaches have been proposed to extend the spectral coverage of these lasers. Recently, we showed that four-wave mixing (FWM) assisted stimulated Raman scattering (SRS) can be harnessed to red-shift high power pulses from 1064 nm to a narrowband output at 1122 nm and 1186 nm and therefore extend the number of accessible fluorophores. In this contribution, we show the applicability of all three wavelengths for multi-photon microscopy and analyze the performance.

## 1. Introduction

Non-linear biomolecular imaging is a growing field with a strong demand for high power lasers. Especially two-photon excitation fluorescence (TPEF) microscopy is an easy-to-use and widely deployed imaging technique [1]. Compared to one photon fluorescence microscopy, TPEF has an intrinsic depth sectioning capability and, together with the deeper penetration compared to one photon excited fluorescence microscopy, it is a real volumetric three dimensional (3D) imaging technique. Compared to other optical 3D imaging modalities, like for example optical coherence tomography (OCT) [2], it has the advantage of good molecular specificity. Fluorescence light is generated only if a fluorophore with absorption corresponding to the excitation wavelength is within the focal volume. Either due to auto-fluorescence or by marking specific sites with a fluorophore, it is ensured that only the molecular structures of interest are visible in the image and the surrounding areas remain dark. Together with its sub-cellular spatial resolution this makes TPEF microscopy a valuable tool for researchers in fields like biomolecular or neuronal science [3-5].

Recent technical advances were made to achieve an even better tissue penetration. Especially imaging in deep cortical layers is of great interest [6] to gain a better understanding of brain functions in mouse models. As scattering and absorption are the main factors limiting tissue penetration in biological samples, the imaging system parameters have to be optimized to reduce these effects to a minimum [7]. Using near infrared (NIR) to extended near infrared (exNIR) light can improve tissue penetration substantially [6-10]. Special red fluorophores have been developed for these applications recently [11-13]. However, high energy pulses are needed to compensate the higher losses for deeper imaging.

All these measures pose strong demands on the excitation light source. The work horse for deeper imaging is currently a combination of a Ti:Sa laser with an optical parametric oscillator [14]. Although this combination has almost ideal excitation parameters, their application is limited to specialized optics labs as they need frequent alignment and maintenance, are expensive, have a large form factor and have strict requirements on the environmental conditions.

To overcome these technical drawbacks, other alternatives are investigated. Amongst different approaches, especially fiber amplified laser diodes are of great interest. They are low cost, inherently fiber based, and very robust [14]. Although femtosecond pulse generation is possible with diode lasers [15], they usually do not achieve as ultra-short pulses as Ti:Sa lasers do. Yet, they have been successfully applied in multi-photon imaging [16, 17] and it has been shown that the overall image quality of picosecond to nanosecond excitation compared to ultra-short pulsed lasers is the same if the same duty cycle is applied [18-20].

The main drawback, however, is that these semiconductor based lasers often lack spectral flexibility as they depend on a secondary gain medium for high pulse energies and these gain media are usually spectrally narrowband. Therefore, the conversion of these high-power pulses to other spectral ranges of particular interest for biomolecular imaging is subject of intensive research [21-26].

In this paper, we demonstrate a simple and straightforward method to extend the spectral coverage of these semiconductor and fiber based lasers for TPEF imaging. Recently, we have shown that a combination of stimulated Raman scattering (SRS) and four-wave-mixing (FWM) in a standard single mode fiber can be harnessed to shift an intense pump pulse to the 1$^{st}$ and 2$^{nd}$ Raman order while keeping the temporal and spectral properties of the pump pulse [27]. We use this FWM seeded SRS wavelength shifting concept to acquire multi-photon images of different stained and unstained samples with a homebuilt laser emitting spectrally narrowband pulses at 1064 nm, 1122 nm, and 1186 nm.

## 2. Methods

*Multi-color laser source*

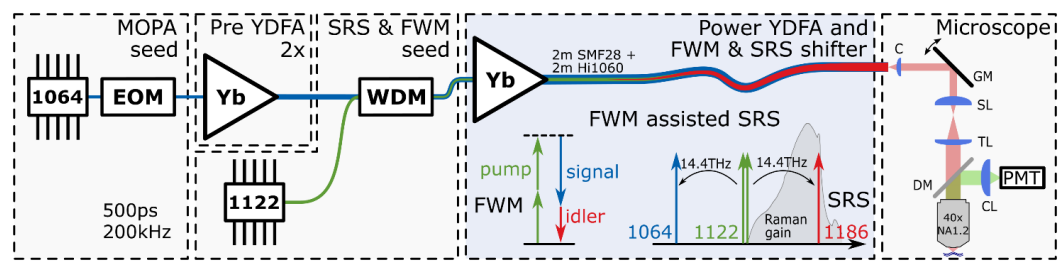

**Fig. 1** Setup of the multi-color master oscillator power amplifier (MOPA). The modulated output of a 1064 nm seed laser diode is amplified in a multistage ytterbium doped fiber amplifier (YDFA) to ~ 1 kW peak power. These high power pulses are shifted to longer wavelengths by a combination of four-wave mixing (FWM) and stimulated Raman scattering (SRS) in passive

standard single mode fiber. By controlling the pump power and the power of the 1122 nm Raman seed laser, the operation can be switched between three wavelengths [27]. The fiber output makes it easy to couple the laser to any multi-photon microscope. EOM: electro optical modulator, Yb: Ytterbium, WDM: wavelength division multiplexer, C: collimator, GM: galvanometric mirrors, SL: scan lens, TL: tube lens, CL: condenser lens, DM: dichroic mirror, PMT: photo-multiplier tube.

The setup of our multi-color fiber laser is depicted in Fig. 1. The details of the setup were described recently [27]. Briefly, we use a master oscillator power amplifier (MOPA) architecture to produce high power pulses which are Raman shifted within the delivery fiber itself. The seed pulses for the MOPA are generated by a narrowband 1064 nm laser diode (LD, Lumics, LU1064M200). The output is actively modulated by a fast 12 GHz electro-optical-modulator (EOM, Photline, NIR-MX-LN-10) which allows adjustable pulse lengths down to 30 ps. The EOM is driven by a homebuilt pulse generator which produces adjustable 500 ps to 10 ns pulses. Faster electronic pulse sources with pulse lengths down to ~30 ps are commercially available and can be implemented in a future setup. The pulse repetition rate can be chosen freely and we typically use rates between 100 kHz-1 MHz. This yields average power values of 10 – 100 mW, comparable to standard femtosecond TPEF imaging [20]. For the following experiments, the pulse length was set to 500 ps and the repetition rate to 200 kHz. This results in a duty cycle of $10^{-4}$ which is about an order of magnitude higher than for ultra-short pulse excitation. Although this is a shortcoming at the moment, implementing a faster pulse source will solve this problem in future. The presented source would be very compatible with fiber beam delivery for endoscopic multi-photon imaging, since, due to the long pulse duration, pulses do not suffer from break up caused by self-phase modulation. Also, due to the spectrally narrow bandwidth of the pulses, chromatic dispersion is no problem either.

The 500 ps seed pulses have a peak power of ~100 mW and are launched into a multistage ytterbium doped fiber amplifier (YDFA). We use two core pumped YDFAs for pre-amplification to peak powers of ~50 W. The final amplification to ~1 kW peak power is accomplished in a multi-mode pumped double-clad (DC) YDFA stage. The DC ytterbium doped fiber (YDF) has a length of 5 m and was chosen to match the mode field diameter of standard single mode fiber which we use for beam delivery. For this delivery fiber, we use a combination of 2 m SMF28 and 2 m Hi1060 fiber, which also acts as non-linear medium for the wavelength shifting. This combination was found to achieve good mode field adaption behind the beam combiner and high nonlinearity for SRS generation [27].

For the wavelength shifting, we insert seed light at 1122 nm before the main power YDFA using a wavelength division multiplexer (WDM) in order to generate stimulated Raman scattering (SRS) in the delivery fiber. Inserting the Raman seed light before instead of after the main power amplifier has two important advantages: First, this measure prevents possible damage to the WDM due to high power pulses. Second, the conversion efficiency reaches values as high as 90 percent [27] by inserting the 1122nm seed before the last amplifier stage, since the Raman threshold can already be reached within the DC-YDFA. Thus, the Raman seed present at this early stage prevents the generation of a broadband Raman plateau and thus makes spectrally narrowband output at 1122 nm possible.

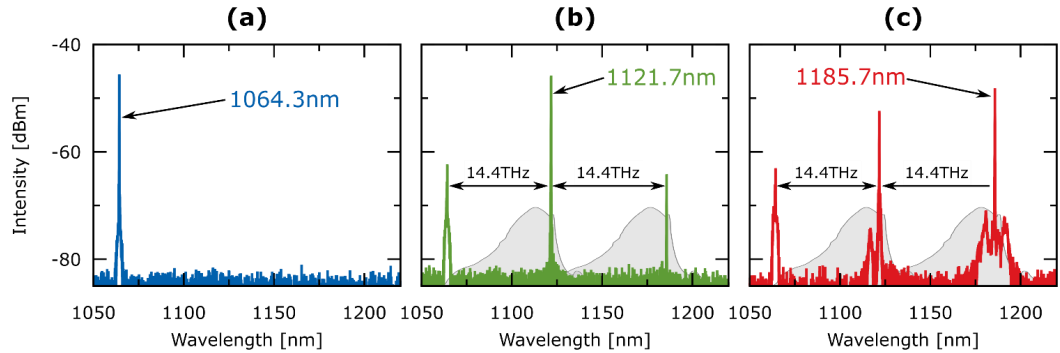

Fig. 2 Emission spectra of the multi-color laser. a) – c) The output spectra when the laser emission is optimized for 1064 nm, 1122 nm, and 1186 nm, respectively. This conversion is the result of the FWM-assisted SRS effect [27]. The Raman gain curve, reproduced to scale after Stolen and Ippen [28], is indicated by grey areas. The spectral components are equidistant in energy with a difference of 14.4 THz as expected from a FWM process.

The wavelength shifting process has been described in detail recently [27]. Basically, we achieve longer wavelengths by red-shifting the intense 1064 nm pulses through stimulated Raman scattering. Compared to other Raman sources [29-32], we insert a seed with 1122 nm at the 1$^{st}$ Raman Stokes order before the main amplifier and we achieve spectrally narrowband high power pulses at this wavelength [31]. As described earlier [33], this also leads to a spectrally narrowband component at 1186 nm, as opposed to the expected broadband Raman gain (cf. figure 2 b) and c), grey area indicates the broadband Raman gain of fused silica, reproduced after Stolen and Ippen [28]). Typical output spectra for the different wavelengths are depicted in figure 2. Figure 2 a), b), and c) show the output spectra when the laser is optimized for operation at 1064 nm, 1122 nm, and 1186 nm, respectively. The three wavelengths are equidistant in energy with a separation of ~14.4 THz, which is consistent with the combination of four-wave mixing (FWM) and stimulated Raman scattering (SRS) in the delivery fiber as underlying processes for the narrowband output at 1186 nm [27]. The high conversion efficiency of up to 90% has two main reasons. As mentioned above, inserting the seed at an early stage reduces a broadband background. The second reason for the high conversion efficiency are the rectangular shaped pulses. As the conversion process is a nonlinear process, intensity changes have a strong impact on the spectral output. With steep edges and a long plateau we achieve a period of low intensity change. With an optimized fiber length and pulse power, almost all of the power can be shifted to longer wavelengths.

In total, we achieve fully electronically switchable, narrowband output pulses at three different wavelengths with maximum peak powers ~1 kW for each wavelength. The fiber-based design, where all wavelengths are delivered by the same single mode fiber, ensures perfect beam quality. This is particularly advantageous in non-linear imaging, where diffraction limited illumination is crucial. Additionally, the fiber enables a flexible microscopy setup where the presented laser source can add multi-photon capability to any fluorescence microscope already present in most optics labs.

*Microscope setup*

We use a home-built multi-photon microscope which follows partly the design presented by Mayrhofer et al. [34]. The light output of the fiber is collimated with an aspheric lens (C, Thorlabs, C280TME-1064) to a beam diameter of ~4 mm ($1/e^2$ intensity drop). We installed a filter wheel directly after the collimator with suitable long-pass or short-pass filters for additional filtering of the excitation wavelengths. Although the out-of-band power for each wavelength is quite low [27], we wanted to filter out remaining light at other wavelengths to avoid unnecessary sample exposure and cross-excitation of other fluorophores. After the filter wheel, a pair of galvanometric mirrors (GM, Thorlabs, GVS002) scans the beam over the sample. Therefore, the pivot point of the galvanometric mirrors is relay imaged to the back focal plane of the microscope objective. We use 2" lenses with long focal lengths to minimize aberrations [35, 36]. A 75 mm achromatic lens (Thorlabs, AC508-075-C-ML) is used as scan lens and a combination of a 300 mm and a 400 mm lens (Thorlabs, AC508-300-C-ML, AC508-400-CML) as tube lens pair. This combination results in a 2.8 fold magnified beam or a beam diameter of ~11 mm filling the back aperture of the objective. With the applied NA 1.2 40x objective (Zeiss, C-Apochromat 40x/1.2 W Corr M27) we achieve an illumination spot size diameter of 560 nm for an excitation wavelength of 1064 nm. The objective has a transmission of ~60 % at the employed wavelengths. Together with the losses of the other components we achieve an overall transmission from the fiber output to the sample plane of ~40 %. The fluorescence light is separated from the excitation light by a dichroic mirror (DM, Thorlabs, DMLP950R) and detected in epi-direction on a fast multi-alkali photomultiplier tube (PMT, Hamamatsu, H12056-20).

*Data acquisition*

In contrast to conventional multi-photon microscopes, the laser and the detection system in our setup are synchronized a priori using an arbitrary waveform generator (AWG). This detection scheme, derived from the TICO-Raman concept [33, 37], allows us to calculate the exact time when a signal is expected, in order to read out the electronics only at that particular point in time. This technique represents a form of time gating which minimizes the noise and thus effectively increases the signal-to-noise ratio (SNR). Apart from electronic noise, also dark current and background light are effectively suppressed by this gating and, furthermore, new measurement modalities like single-pulse fluorescence lifetime measurements (SP-FLIM) are possible [38].

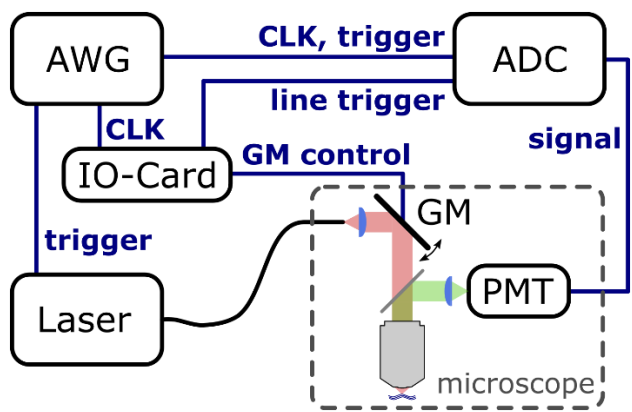

**Fig. 3** Electronic synchronization of the homebuilt multi-photon microscope. An arbitrary waveform generator (AWG) provides a sample clock (CLK) reference signal and trigger signals for all components. ADC: analog-to-digital converter card, PMT: photomultiplier tube, GM: galvanometric mirrors.

A schematic setup of our synchronization is depicted in figure 3. An arbitrary waveform generator (AWG, TTi, TGA12104) provides the necessary timing signals. One channel is used for triggering the laser. Another channel provides a sample clock (CLK) for a digital input-output card (IO-Card, National Instruments, NI-PCI-6285) which controls the galvanometric mirrors (GM) and also outputs a line trigger signal to the analog-to-digital converter card (ADC). The ADC is fed an external sample clock (CLK) and a trigger signal from the AWG. This leads to a complete electronic synchronization, allowing calculating the exact time when a fluorescence signal is expected on the PMT.

Before the digitizer, the current output of the PMT is converted to a voltage by a 50 Ohm termination directly after the PMT. This ensures a low time constant as the capacity of a connected BNC cable can then be neglected. Together with a 50 Ohm terminated input of the analog-to-digital converter card (ADC, Alazartech, ATS9373) we achieve a total current to voltage conversion of 25 V/A. The ADC card has a 2 GHz analog bandwidth, a sampling rate of 4 GS/s and is capable of continuously streaming the data directly into the PC's memory. This is useful if long time traces of non-repetitive events are to be recorded. In order to determine the expected signal of a single photon detection, we have to consider first the gain of the PMT. Typically, we set the control voltage of the PMT to ~800 mV which results in an internal gain of $g \sim 3 \cdot 10^5$. Thus, at a gating time of $\tau \sim 1$ ns, a single photon event produces a current of $I = e \cdot g / \tau \sim 48$ µA resulting in a voltage amplitude of $U = R \cdot I \sim 1.2$ mV for a single photon detected in the ADC. Assuming a linear response of the PMT, the full range of ±400 mV of the ADC corresponds to a total of ~330 detectable photons per time gate. As the rms input noise of the ADC is less than 0.5 mV, even a single photon event has an SNR bigger than unity and so no amplifier is required which could potentially induce electronic distortions on the signal and deteriorate the SNR through the added electronic amplifier noise.

*Sample preparation*

As an example for the applicability of our multi-wavelength laser we acquired images with three different configurations for the three wavelengths 1064 nm, 1122 nm, and 1186 nm, respectively. First, we acquired images with 1064 nm excitation. As a first sample we used a slice of a *convalaria majalis* stem stained with acridine orange (Lieder GmbH). Then, to validate that the excitation power is sufficient for autoflourescence imaging, we recorded a TPEF-image of a *ficus benjamina* plant leave. To demonstrate second harmonic generation (SHG) imaging in biological tissue, we acquired an image of a mouse ear. For mouse tissue collection C57Bl/6J mice were sacrificed and skin samples were prepared. This procedure was approved by the Animal Care and Use Committee (Kiel, Germany: 5_2016-01-19) and was performed by certified personnel.

Second, we show multi-color excitation TPEF images acquired with the red shifted 1122 nm and 1186 nm laser wavelengths. For this, we used COS-7 cells stained with two red fluorophores. The microscope slide was kindly provided by Leica Microsystems. Within the cells, tubulin is labeled with Star-635 (Abberior) and mitochondria with Atto-594 (ATTO-TEC). Figure 4 shows the absorption and emission spectra of both fluorophores for one-photon excitation. The excitation wavelengths at 1122 nm and 1186 nm are shown at half their wavelength for comparison. Although not perfectly coinciding with the maximum absorption peaks, excitation at 1122 nm and 1186 nm should be suited well for multi-color excitation of the two fluorophores. The presented one-photon absorption spectra represent only a guideline for the two-photon absorption spectra, since differences can arise from different quantum-mechanical selection rules for the absorption processes. For many fluorophores it was found that exciting at a wavelength corresponding to the typical blue-shifted shoulder of the one-photon absorption spectrum exhibits the same or even a stronger two-photon absorption than exciting at the one-photon absorption peak [11, 39].

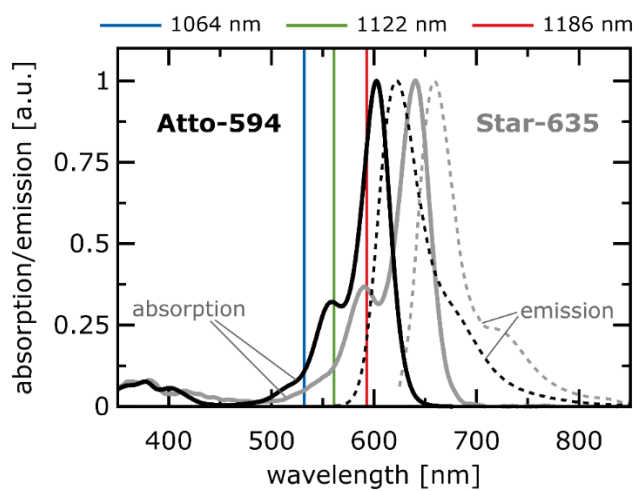

**Fig. 4** One-photon absorption and emission spectra for the dyes Atto-594 and Star-635 which are used to label the mitochondria and tubulin in the COS-7, respectively. Solid curves are absorption spectra, dashed curves emission spectra. Colored lines indicate half values of the available excitation wavelengths.

It should be noted that both fluorophores have strong absorption at 1186 nm. However, at 1122 nm predominantly Atto-594 is excited (cf. fig. 4). Therefore, we expect to see both structures in the 1186 nm excitation channel and mainly the mitochondria in the 1122 nm excitation channel. In fluorescence emission, both fluorophores could be distinguished by spectrally selective detection. Here, we employ simple image processing instead to visualize the two individual components.

## 3. Results & Discussion

For TPEF imaging, the laser parameters were set to a pulse length of 500 ps and a repetition rate of 200 kHz. The images at 1064 nm excitation are depicted in figure 5. These images have a size of 1024x1024 pixel and cover a region of 450 µm x 450 µm. They were acquired using ~25 mW excitation power on the sample and a pixel dwell time of 20 µs. Figure 5 a) shows the TPEF signal of an acridine orange stained *convalaria majalis* stem and b) of an unstained *ficus benjamina* plant leave,. Figure 5 c) shows the SHG signal of collagen and elastin within a mouse ear. With the applied excitation parameters, no degradation of the sample was observed.

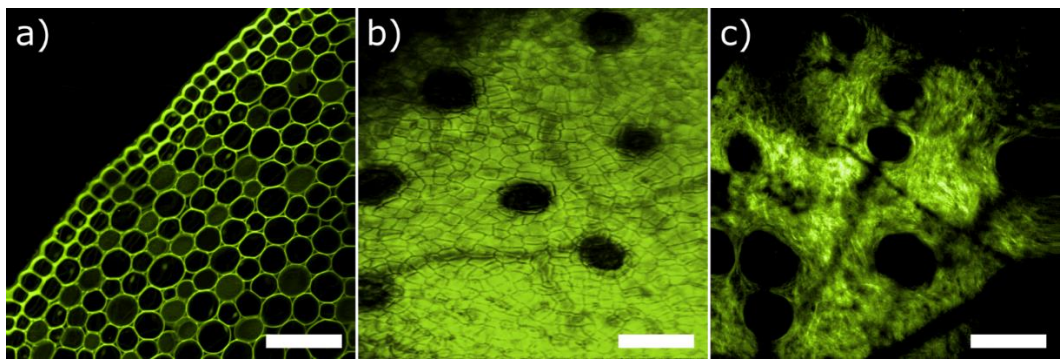

Fig. 5 Imaging with 1064 nm excitation light. a) TPEF image of a stained slice of *convalaria majalis*. b) TPEF auto fluorescence image of a *ficus benjamina* plant leave. c) SHG image of mouse skin. Scale bar 100 µm.

For imaging with the longer wavelengths at 1122 nm and 1186 nm, the Raman seed laser diode at 1122 nm was switched on and the pump power of the DC-YDFA was adjusted for optimal performance for each individual wavelength. After collimation, the remaining out-of-band light was filtered out by a combination of a long-pass and a short-pass filter. The power on the sample was set to 30 mW for each wavelength. Figure 6 shows the TPEF images of the stained COS-7 cells. They were imaged sequentially with 1122 nm light first, followed by excitation at 1186 nm. Both images have a size of 1024x1024 pixels, cover a region of 180 µm x 180 µm and were recorded with a pixel dwell time of 100 µs. This relatively long pixel dwell time is partly due to the weak signal of the sample and partly due to low repetition rate which was chosen for a small duty cycle. If we would reduce our pulse duration to 50 picoseconds, we would be able to increase the repetition rate to 2 MHz while keeping the CW power constant.

Figure 6 a) shows the COS-7 cells excited at 1122 nm. Here, only Atto-594 is efficiently excited which visualizes the mitochondria. Figure 6 b) was acquired at 1186 nm excitation which excites both Atto-594 and STAR-635 fluorophores. Therefore, both structures, the mitochondria and the tubulin are visible in this 1186 nm channel. To distinguish both structures, we applied a simple image processing to enhance the difference in both images. For visualizing the tubulin, we subtracted the image acquired with 1122 nm excitation from the image acquired with 1186 nm which subtracts the mitochondria signal and thus effectively isolates the tubulin structure (Fig. 6d). An overlay of this processed and blue colored image with the green colored image acquired with 1122 nm excitation can be seen in figure 6e). There, the tubulin structures in blue can be clearly distinguished from the mitochondria structures in green.

In summary, these two simple steps, the multi-wavelength excitation at 1122 nm and 1186 nm and the following digital extraction process help visualizing both fluorophores, and hence, the targeted structures very well without the need for a spectrally selective detection.

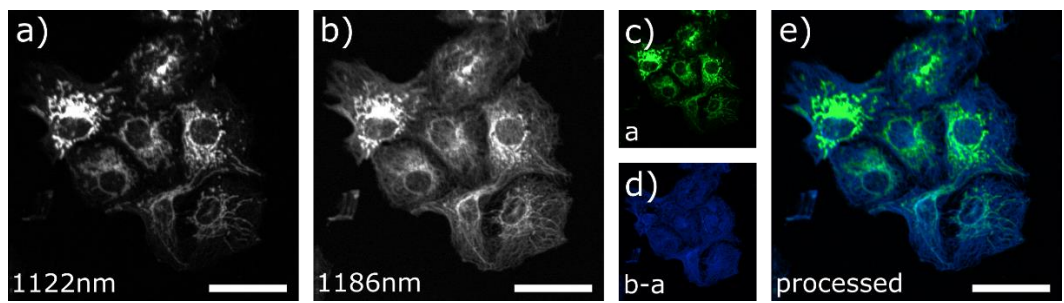

Fig. 6 Two-photon imaging of stained COS-7 cells with long wavelength excitation. a) 1122 nm excitation channel. b) 1186 nm excitation channel. Images c) and d) are the color-coded fluorochrome channels, where d) is a result of subtracting a) from b) to isolate the Atto-594 channel. e) Processed overlay, where green color shows mitochondria (Atto-594) and blue color tubulin (STAR-635). Scale bar 50 µm.

## 4. Conclusion and Outlook

In this paper, we presented a simple and straightforward way to use FWM seeded SRS for extending the spectral coverage of single wavelength fiber lasers. Our actively modulated 1064 nm fiber MOPA provides high power pulses for frequency conversion in the delivery fiber. By carefully controlling the seed and pump powers, we can select three different excitation wavelengths. We presented TPEF imaging of plants and SHG imaging of mouse skin at 1064 nm excitation. The applicability of the longer wavelengths 1122 nm and 1186 nm was shown by imaging COS-7 cells labeled with red fluorophores. The combination of long pulse durations and longer wavelengths promises to be ideal for deep tissue imaging. Furthermore, the different excitation wavelengths can be switched at very high speed. We previously showed that the FWM seeded SRS wavelength conversion method is capable of switching the wavelength in a pulse-to-pulse manner. This high-speed capability can prove useful in biological and medical hyperspectral fluorescence imaging. We showed that with this technique, two fluorophores can be distinguished merely by multi-wavelength excitation, no spectrally selective detection is needed. A second detector can thus be omitted which can be a great advantage if the system has to be miniaturized for example for a handheld imaging device. Future work will be focused on developing such a handheld or endoscopic multi-color multi-photon device which also harnesses the flexible fiber beam delivery of the presented laser system.

**Funding**

This Research was funded by the European Union project ENCOMOLE-2i (Horizon 2020, ERC CoG no. 646669), the German Research Foundation (DFG project HU1006/6 and EXC 306/2), the European Union within Interreg Deutschland-Danmark from the European Regional Development Fund in the project CELLTOM, the German Federal Ministry of Education and Research (BMBF no. 13GW0227B: "Neuro-OCT") and the Schleswig-Holstein Excellence-Chair program within the Cluster of Excellence Inflammation at Interfaces jointly funded by the German Research Foundation (DFG) and the state of Schleswig-Holstein.

**Acknowledgments**

The authors thank Nadine Merg from the Lübeck Institute of Experimental Dermatology (LIED) for preparing the mouse samples.

**Disclosures**

The authors declare that there are no conflicts of interest related to this article.